\documentstyle[12pt,epsf]{article}

\newcommand{\eq}[1]{Eq.~(\ref{#1})}

\def\bea{\begin{eqnarray}}
\def\eea{\end{eqnarray}}

\def\be{\begin{equation}}
\def\ee{\end{equation}}

\def \hy{{\hat y}}
\def \tR{{\tilde R}}
\def \tw{{\tilde \omega}}

\begin{document}
\begin{titlepage}

\hfill\vbox{\normalsize\hbox{hep-th/0202049}
    \hbox{YITP-SB-02-05}
    \hbox{February, 2002}}
  \bigskip

\begin{center}
{  \large\sf Brane-Bulk Interaction in Topological Theory.}
\end{center}
\date{\small\sf\today}
\pagestyle{empty}
\centerline{A.~Boyarsky
\footnote{e-mail:boyarsky@alf.nbi.dk}
\footnote{On leave from Bogolyubov ITP, Kiev , Ukraine} }
\smallskip
\centerline{\it Niels Bohr Institute }
\centerline{\it Blegdamsvej 17, DK-2100 Copenhagen , Denmark}
\bigskip
\centerline{B.~Kulik
\footnote{e-mail:bkulik@insti.physics.sunysb.edu}}
\smallskip
\centerline{\it C.N. Yang Institute for Theoretical Physics}
\centerline{\it State University of New York}
\centerline{\it Stony Brook, N. Y. 11794-3840, U.S.A.}
\smallskip

\baselineskip 18pt


\begin{abstract}
In this letter we address the problem of inducing
boundary degrees of freedom from a bulk theory
whose action contains higher-derivative corrections.
As a model example we consider a topological theory with
an action
that has only a ``higher-derivative'' term. By choosing
specific coupling of the brane to the bulk we
show that the boundary action contains gravity action
along with some higher-derivative corrections.
The co-dimension of the brane is more than one. In this sense
the boundary is singular. 
\end{abstract}

\end{titlepage}
\newpage
\pagestyle{plain}
\setcounter{page}{1}

\section{Introduction}

The subject of inducing boundary degrees of freedom from the
bulk has a very rich history.
There exists a rather general mechanism for
gauge degrees of freedom in a topological theory
to become dynamical after introduction of a boundary
(see e.g.\cite{Seiberg,Kogan-Carlip,banados}) .


In this paper we consider a similar mechanism but with
one unusual ingredient: the boundary 
of the manifold will be singular.
It is ``singular'' in the sense that its co-dimension is more than one
(e.g. marked four dimensional sub-manifold embedded in six dimensions).
To treat the sub-manifold as a regular boundary of co-dimension one
we ``regularize it''.
We blow it up to a cylinder and then work with the boundary of this cylinder.
At the end we take the limit in which the cylinder shrinks
back to the original singular sub-manifold.

The motivation to study singular boundaries comes from the problem
of studying the dynamics of solitonic (brane) backgrounds.
One often uses the approximation in which 
the theory of localized zero-modes is separated from the rest of the 
bulk modes.
Instead of original theory one considers  the theory on
trivial (e.g. flat) background in the bulk plus the
lower-dimensional theory in
the world-volume of the brane as a theory of localized zero modes of the brane.
The total action of the system is a sum of two actions corresponding
to each theory. In such an 
approximation there is a question of how an interaction
between the brane and the bulk should be taken into account.
One possible regime is when both theories decouple from each other.
However, sometimes it is impossible to neglect the interaction.
E.g. if the dimension of the sub-manifold (brane's world-volume) is even
and some of zero modes are chiral, the world volume theory could suffer from 
gauge and gravitational anomalies. In this case the world-volume theory
is inconsistent by itself (which actually means that it is the decoupling approximation
which is inconsistent ) and it is necessary to take into account
an interaction between bulk and world-volume which makes the whole 
theory anomaly free. This is called ``inflow mechanism''\cite{ch}  of 
anomaly cancellation.
There are several important examples of such  cancellation in field and M theory
\cite{fhmm,bk,Bonora,jh-or}. In general the bulk theory has non-zero gauge variation 
which is non-zero only on the sub-manifold and cancels the anomalous gauge
variation of the world-volume theory.

In this paper 
we study some other example of such an interaction though we use the same setup.
The bulk action is purely topological. 
Topological terms are not unusual for string theory.
Some of M-theory corrections to 11-d SUGRA action have
structure of lower dimensional topological terms embedded in eleven
dimensions \cite{Tseytlin}. In general such terms can appear
as a higher-derivative correction to some more complicated system.
These corrections can have very different origin depending on the bulk
theory. For simplicity we consider only topological term 
by itself to study a new interaction it can be responsible for. 

The interaction with the brane is specified by choosing
boundary conditions for the bulk fields.
Our goal will be to show that introduction of the boundary in
this topological theory under some rather general
boundary conditions
generates  the boundary term that contains lower dimensional gravity
action.

This work generalizes the result of \cite{Bo-Bo}. That paper
gave the realization to the idea suggested by 't Hooft
of canceling the four dimensional cosmological constant
by inducing the gravity from topological six dimensional theory.
Here, without any relation to cosmological constant problem, 
 we give more accurate mathematical formulation
of the mechanism of inducing gravity on the brane from topological term
in the bulk. 
We generalize the construction to the case of higher co-dimensions
which can arise in other applications. This generalization is not quite
trivial since in the case of co-dimension higher than two
the angular form has more complicated dependence on the normal bundle
gauge connection.

Besides, we make one more improvement of the construction used in 
\cite{Bo-Bo}. In this work the boundary conditions were used which
violated the covariance under rotations of normal bundle.
In this work we resolve that difficulty  by finding suitable boundary conditions
that are covariant under normal bundle rotations. These boundary conditions 
have a natural physical interpretation. 

One of the interesting implications of our result is in the context of 
brane-worlds. It can offer mechanism of localizing gravity. We
will give a short discussion of that in the conclusion.

\section{Embedding}
Consider 4-d sub-manifold embedded into six dimension, $W^4\subset M^{6}$. 
This embedding is
``singular'' in a sense that it can't be treated as a boundary because
the boundary of 6-d manifold is 5 dimensional. To ``regularize ``
such embedding it is convenient to introduce tabular neighborhood 
$W_\epsilon$ of 4-d sub-manifold \cite{Bott-Tu}. Locally 
$ W_\epsilon \times D_\epsilon^2 $ , where $D_\epsilon^2$ is a 2-d
disk of radius $\epsilon$. That is, tabular neighborhood  is a cylinder 
surrounding the brane.
The theory in the bulk is defined on the six dimensional manifold
with the tabular neighborhood cut out. Its boundary is the boundary
of the $\partial W_\epsilon$ of the tabular neighborhood. 
Introduction of the boundary requires to
impose some boundary conditions for the bulk theory. This will specify 
an interaction between bulk and brane theories. 

\section{The action}

For simplicity the action we want to consider is purely topological. 
That is, it can be locally expressed as a total derivative.
In terms of forms it is written as
\bea
E_6=\int_{M^6} \varepsilon_{A B C D E F} \tR^{A B} \wedge 
\tR^{C D}\wedge \tR^{E F}
\eea

Where $\tR$ is a curvature of Lorenz spin-connection. Thus $E_6$ represents
an Euler class. Since the manifold $M^6$ has a boundary
$\partial W_\epsilon$ such an action can be written
as a surface term only. We will proceed with determining it.

On the brane the original $SO(1,5)$ Lorenz group is broken to 
$SO(1,3) \times SO(2)$. Let's split the 6-d Lorenz connection
$\tw^{AB}$ into corresponding parts:
\bea
\tw^{ab}  = A^{ab} ~~~~  
\tw^{a\alpha} = \pi^{a\alpha} ~~~~~ 
\tw^{\alpha\beta} = \omega^{\alpha\beta} \label{con}
\eea
where $a$,$b$ are the indexes in the $SO(2)$ part of the bundle
and $\alpha$, $\beta$ in the $SO(1,3)$.
In these terms the 6-dim curvature
$\tR^{A B} = d \tw^{A B} + \tw^{A C} \wedge \tw_C^{~B}$ is :
\bea
\tR^{ab} &=& F^{ab}(A) - \pi^a_\alpha \wedge \pi^b_\alpha \\
\tR^{a\alpha} &=& D(A,\omega)~\pi^{a\alpha} \\
\tR^{\alpha\beta} &=& R^{\alpha\beta}(\omega) - 
\pi^{\alpha}_a \wedge \pi^{\beta}_a \label{cur}
\eea \noindent
Where $F^{a b}(A)$ and $R^{\alpha\beta}(\omega)$ are curvatures
that correspond to connections $A$ and $\omega$ ,
$D(A,\omega)~\pi^{a\alpha} $ is a covariant derivative with respect
to both bundles $SO(2)$ and $SO(1,3)$
\bea
D(A,\omega)~\pi^{a\alpha} = d \pi^{a\alpha} + 
{\omega^\alpha}_\beta \wedge \pi^{a\beta} +
{A^a}_b \wedge \pi^{b\alpha}  
\eea 
Now we can express the Euler form as
\bea
E_6 = \int_{M^6} \varepsilon_{ab} \varepsilon_{\alpha\beta\gamma\delta}~
d\Big[ 3 R^{\alpha\beta}\wedge R^{\gamma\delta}\wedge A^{a b}
- 6 \pi^{a\alpha} \wedge D\pi^{b\beta} \wedge 
( 2 R^{\gamma\delta} - \pi^\gamma_c \wedge \pi^\delta_c ) 
\Big] = \\
\oint_{\partial W_\epsilon} 
\varepsilon_{ab} \varepsilon_{\alpha\beta\gamma\delta}~
\Big[ 3 R^{\alpha\beta}\wedge R^{\gamma\delta}\wedge A^{a b}
- 6 \pi^{a\alpha} \wedge D\pi^{b\beta} \wedge 
( 2 R^{\gamma\delta} - \pi^\gamma_c \wedge \pi^\delta_c ) 
\Big] \label{BA}
\eea

\section{Angular form}

The boundary term we just obtained contains the integration 
over the  whole boundary $W^4\times S_1$.
We would like to reduce it to integration over $W^4$ only by performing
integration over $S^1$ separately. In doing so we define first the 
form integration of which over the transverse directions is equal to one.
This is the volume form. Let us introduce the coordinates on a unit sphere
$S^1$, $\hy^a = y^a/y$. Then the volume form can be expressed as
\bea
\Psi_1 = \frac{1}{2\pi} \varepsilon_{ab} \hy^a d\hy^b
\eea
The boundary of the tabular neighborhood is isomorphic to the total space
of the $SO(2)$ bundle, normal bundle. The base of the normal bundle is
sub-manifold $W^4$ and $S^1$ are fibers. 
 Since we want to perform an integration along the fiber
we  need to  introduce covariant
generalization of the volume form which will be globally defined \cite{Bott-Tu}
It requires an introduction of the
connection on the normal bundle. The resulting form $e_1$ is called an angular
form  and has the following properties. 
Its restriction on the fibers is a volume form and
\bea
d e_1 = \chi(F)
\eea
Where $\chi(F)$ is an Euler class of the normal bundle.
Such angular forms can be constructed for the case of any even co-dimension.
In the case of odd co-dimension the corresponding angular form is closed.
We consider the cases of $e_1$ . The explicit expression is
\bea
e_1 = \frac{1}{2\pi} \varepsilon_{ab} \hy^a D \hy^b \label{e1}
\eea
Where $D\hy^a = d\hy^a + {\Theta^a}_b \hy^b$, $\Theta$ is a connection
on the normal bundle.

\section{Boundary conditions}
Next we impose some boundary conditions on the 
connection
${\tilde \omega}_a^{~b}$, that is, specify the coupling of the bulk 
theory to the brane. We want to do it in such a way that the boundary
action \eq{BA} splits into the product of
two parts, the angular form and the rest that depends only on the brane 
coordinates. Then we can perform the integration and get the action
defined on the brane only.
First, it is required that :
\bea
\pi^{a\alpha} \mid_{\partial W_\epsilon} = \hy^a e^\alpha \label{veilb}
\eea \noindent
Next, we require $e^\alpha$ to depend on brane coordinates only
and to satisfy no-torsion constrain with respect to connection
$\omega^{\alpha\beta}$, that is $D(\omega) e = 0$.
Later we'll see that $e^\alpha$ plays a role of induced veilbein on the brane. 
Under such conditions 
\bea
\pi^{a\alpha} D(\omega,A)\pi^{b\beta} = - \hy^a D(A)\hy^b \wedge 
e^\alpha \wedge e^\beta 
\eea
Second, part of the connection $A$ is taken as the connection on the normal
bundle $\Theta$ , the rest will yield the angular form $e_1$.
\bea
A^{a b} \mid_{\partial W_\epsilon} = 
a \hy^{[a} D(\Theta) \hy^{b]} + \Theta^{a b}
\eea \noindent
Where brackets stand for anti-symmetrization.
With such a choice of the connection $\Theta$ , $D(A) \hy^a$ is
\bea
D(A) \hy^a = D(\Theta)\hy^a + a \hy^{[a} D(\Theta) \hy^{b]} \hy^b =
(1-a) D (\Theta)\hy^a
\eea
Since $\hy^a \hy^a = 1$ and $\hy^a D \hy^a = 0$.

Here we should make a short remark. It
may look that chosen boundary conditions are very artificial.
Nevertheless, one can show though that they
are  a consequence a very simple requirement of spherical
symmetry. All modes which are spherically symmetric
in the plane normal to the brane satisfy them.

\section{Boundary action}

Now we can calculate the boundary action \eq{BA} under chosen
above boundary conditions.
\bea
E_6 &=&
\oint_{\partial W_\epsilon} 
\varepsilon_{a b} \varepsilon_{\alpha\beta\gamma\delta}~
6 \hy^a D \hy^b \nonumber \\
&\wedge& \Big[ a R^{\alpha\beta} \wedge R^{\gamma\delta} 
+ 2(1-a) R^{\alpha\beta} \wedge e^\gamma \wedge e^\delta
- (1-a) e^\alpha \wedge e^\beta \wedge e^\gamma \wedge e^\delta
\Big] \nonumber \\
&+& 
\oint_{\partial W_\epsilon} 
\varepsilon_{a b} \varepsilon_{\alpha\beta\gamma\delta}~
3 \Theta^{a b}\wedge  R^{\alpha\beta} \wedge R^{\gamma\delta}
\eea
Thus we succeeded in separating angular form and fields on the brane.
The $\Theta$-dependent term doesn't contribute since the integrand form
$\Theta RR$ doesn't have any transverse components.
We can perform integration of the rest to get
\bea
E_6 &=&
\oint_{\partial W_\epsilon} 12\pi e_1 \nonumber \\
&\wedge& \varepsilon_{\alpha\beta\gamma\delta}
\Big[ a R^{\alpha\beta} \wedge R^{\gamma\delta} 
+ 2(1-a) R^{\alpha\beta} \wedge e^\gamma \wedge e^\delta
- (1-a) e^\alpha \wedge e^\beta \wedge e^\gamma \wedge e^\delta
\Big] \\
&=&
12 \varepsilon_{\alpha\beta\gamma\delta}~\int_{W^4} \nonumber \\
&a& R^{\alpha\beta} \wedge R^{\gamma\delta} 
+ 2(1-a) R^{\alpha\beta} \wedge e^\gamma \wedge e^\delta
- (1-a) e^\alpha \wedge e^\beta \wedge e^\gamma \wedge e^\delta
\eea 

Thus the $E_6$ term is equivalent 
to the following action on the brane: a topological term, 
Hilbert-Einstein action and a cosmological term.

\section{ Generalization to co-dimension 4}

The whole frame work can be easily generalized to the cases
of higher co-dimension. We consider the case of co-dimension four.
Thus we have four dimensional sub-manifold $W^4$ embedded into
eight dimensions. The action is taken to be eight dimensional
Euler class
\bea
E_8=\int_{M^8} \varepsilon_{A B C D E F G H} \tR^{A B}\wedge \tR^{C D}\wedge 
\tR^{E F} \wedge \tR^{G H}
\eea

On the brane the original $SO(1,7)$ Lorenz group is broken to
$SO(1,3) \times SO(4)$. Splitting of the 8-dimensional spin-connection
$\tw$ and the curvature tensor $\tR$ stays the same as in \eq{con} and \eq{cur}
correspondingly except that index $a$ in $A=(\alpha,a)$ is in $SO(4)$
group now. Since $E_8$ is a closed form it can be written locally as a total
derivative. In terms of the decomposition of $SO(1,7)$ fields into
$SO(1,3) \times SO(4)$ fields it reads
\bea
E_8 &=& \int_{M^8} \varepsilon_{a b c d}~\varepsilon_{\alpha\beta\gamma\delta}~
d\Big[ 6 R^{\alpha\beta}\wedge R^{\gamma\delta}\wedge 
CS(A)^{a b c d} \nonumber \\
&+& \pi^{a\alpha} \wedge D\pi^{b\beta} \wedge 
\Big( 
16 D\pi^{c\gamma} \wedge D\pi^{d\delta}
+ 24 ( R^{\gamma\delta}\wedge \phi^{c d} + 
F^{c d} \psi^{\gamma\delta}) \nonumber\\
&-& 48 R^{\gamma\delta}\wedge F^{c d} 
- 16 \phi^{c d} \wedge \psi^{\gamma\delta}
\Big) 
\Big] = \\ 
&=& 
\oint_{\partial W_\epsilon} 
\varepsilon_{a b c d}~\varepsilon_{\alpha\beta\gamma\delta}~
\Big[ 6 R^{\alpha\beta}\wedge R^{\gamma\delta}\wedge 
CS(A)^{a b c d} \nonumber \\
&+& \pi^{a\alpha} \wedge D\pi^{b\beta} \wedge 
\Big( 
16 D\pi^{c\gamma} \wedge D\pi^{d\delta}
+ 24 ( R^{\gamma\delta}\wedge \phi^{c d} + 
F^{c d} \psi^{\gamma\delta}) \nonumber \\
&-& 48 R^{\gamma\delta}\wedge F^{c d} 
- 16 \phi^{c d} \wedge \psi^{\gamma\delta}
\Big) 
\Big] \label{BA8}
\eea
Where $CS(A)$ is Chern-Simons form of the $SO(4)$ connection $A^{a b}$
\bea 
CS(A)^{abcd} =
d A^{a b} \wedge A^{c d} + \frac{2}{3} A^{a x} \wedge {A_x}^b \wedge A^{c d}
\eea  
and $\phi^{a b}$ and $\psi^{\alpha\beta}$ are defined as
\bea
\phi^{a b} = \pi^a_\gamma \wedge \pi^b_\gamma ~~~~~~~~~
\psi^{\alpha\beta} = \pi^\alpha_c \wedge \pi^\beta_c 
\eea 
Before introducing the boundary conditions we want to discuss 
the angular form in this case. Its explicit expression is
\bea
e_3 = \frac{1}{2\pi^2} \varepsilon_{a b c d}
\Big[
\frac{1}{2} \hy^a D \hy^b \wedge D \hy^c \wedge D \hy^d -
\frac{1}{3} \hy^a F(\Theta)^{b c} \wedge \hy^d
\Big] \label{e3}\\
d e_3/2 = \chi(F) =  \frac{1}{32\pi^2} \varepsilon_{a b c d} 
F(\Theta)^{ab}\wedge F(\Theta)^{cd} \label{de3}
\eea
Where the covariant derivative is taken with respect to the connection
on the normal bundle $\Theta$.
The first term in $e_3$ contains the volume form on $SO(4)$ ,
$\Psi_4 = \varepsilon_{abcd}~\hy^a d\hy^b\wedge d\hy^c\wedge d\hy^d$,
the rest is required by the condition of \eq{de3}.

The boundary conditions in this case are very similar to the case of lower
co-dimension.
\bea
\pi^{a\alpha} \mid_{\partial W_\epsilon} = \hy^a e^\alpha
\eea \noindent
That implies the following for $\phi^{ab}$ and $\psi^{\alpha\beta}$
\bea
\phi^{ab} \mid_{\partial W_\epsilon} = 0
~~~~~~~~~~~~
\psi^{\alpha\beta} \mid_{\partial W_\epsilon} = e^\alpha\wedge e^\beta
\eea
Next, we require $e^\alpha$ to depend on brane coordinates only
and to satisfy no-torsion constrain with respect to connection
$\omega^{\alpha\beta}$, that is $D(\omega) e = 0$
Under such conditions 
\bea
\pi^{a\alpha} D(\omega,A)\pi^{b\beta} = - \hy^a D(A)\hy^b \wedge 
e^\alpha \wedge e^\beta 
\eea
Second, part of the connection $A$ is taken as the connection on the normal
bundle $\Theta$
\bea
A^{a b} \mid_{\partial W_\epsilon} = 
a \hy^{[a} D(\Theta) \hy^{b]} + \Theta^{a b}
\eea \noindent
The term $\varepsilon_{abcd} \hy^a D(\theta) \hy^b F(A)^{cd}$ will
give the angular form $e_3$
\bea
F(A)^{cd}  \mid_{\partial W_\epsilon}
= F(\Theta)^{cd} + a \hy^{[c} F(\Theta)^{d]x} \hy^x
+ a(2-a) D(\Theta)\hy^c \wedge D(\Theta)\hy^d \\
\varepsilon_{abcd} \hy^a D(\theta) \hy^b F(A)^{cd} =
\varepsilon_{abcd} \hy^a D(\theta) \hy^b \wedge
\Big(
 F(\Theta)^{cd} + a(2-a) D(\Theta)\hy^c \wedge D(\Theta)\hy^d
\Big)
\eea
With such choice of the connection $\Theta$ , $D(A) \hy^a$ is
\bea
D(A) \hy^a = (1-a) D (\Theta)\hy^a
\eea
And thus the term $\pi D\pi D\pi D\pi$ gives
\bea
\pi^{a\alpha} D(\omega,A) \pi^{b\beta}\wedge D\pi^{c\gamma}\wedge 
D\pi^{d\delta}  \mid_{\partial W_\epsilon}
= \\
= (1-a)^3 \hy^a\wedge D(\Theta)\hy^b\wedge D\hy^c\wedge D\hy^d  \wedge
e^\alpha\wedge e^\beta\wedge e^\gamma\wedge e^\delta
\eea
Now we are ready to compute boundary action \eq{BA8}.
\bea
E_8  &=& \oint_{\partial W_\epsilon} 
\varepsilon_{\alpha\beta\gamma\delta}~
\Big[ (12a + 4a^3) e_3\wedge R^{\alpha\beta}\wedge R^{\gamma\delta}
\nonumber \\
&+& 16 (1-a)^3 e_3\wedge e^\alpha\wedge e^\beta\wedge e^\gamma\wedge e^\delta
- 24 (1-a) e_3\wedge e^\alpha \wedge e^\beta \wedge e^\gamma\wedge e^\delta 
\nonumber \\
&+& 48 (1-a) e_3\wedge R^{\alpha\beta}\wedge e^\gamma \wedge e^\delta 
\Big] + \Phi(\Theta,R,e,\hy)
\eea
Where $\Phi(\Theta,R,e,\hy)$ represents all terms that do not have enough
components in transverse directions to contain the volume form.
Thus $\oint \Phi$ = 0.
Now we can perform the integration to get
\bea
E_8  &=& \int_{W^4} 
\varepsilon_{\alpha\beta\gamma\delta}~
\Big[ (12a + 4a^3) R^{\alpha\beta}\wedge R^{\gamma\delta}
\nonumber \\
&+& \Big( 16 (1-a)^3 - 24 (1-a) \Big) 
e^\alpha \wedge e^\beta \wedge e^\gamma \wedge e^\delta \nonumber \\
&+& 48 (1-a) R^{\alpha\beta}\wedge e^\gamma \wedge e^\delta 
\Big]
\eea
We once again see that $E_8$ term with the set above boundary conditions
yields a topological, a cosmological and Hilbert-Einstein terms
on the brane.


\section{ Conclusions and discussion.}
\label{concl}

In this letter we addressed the problem of inducing
boundary degrees of freedom from a bulk theory
whose action contains higher-derivative corrections.
As a model example we considered a topological theory with an
action that has only a ``higher-derivative'' term. By choosing
specific coupling of the brane to the bulk we
showed that the boundary action contains gravity action
along with some higher-derivative corrections.
The co-dimension of the brane is more than one. In this sense
the boundary was singular. 

This result is refinement and generalization of the work done in \cite{Bo-Bo}.
First of all we considered the case of higher co-dimension.
The non-trivial part of it lies in the difference between
$e_1$ and $e_3$ forms , \eq{e1},\eq{e3}. The normal bundle connection
enters $e_1$ in a straightforward way , it just make the volume form covariant.
On the other hand $e_3$ is the first non-trivial case when an angular
form contains other terms besides a covariantized volume form.

There is another (more important) new result. The boundary conditions
in \cite{Bo-Bo} broke the covariance under rotations in the normal bundle.
The analog of the condition in \eq{veilb} was
that only one component $\pi$ contained 4-dimensional veilbein,
the other was set to zero. That corresponded to choosing one fixed
normal vector out of all normal vectors. In this work (as \eq{veilb} shows)
we keep the normal vector arbitrary and integrate over all of them in 
the action. In this way the covariance with respect to rotations in the
normal bundle is preserved by the boundary conditions.
The topological theory considered doesn't have any metric in the bulk.
It has only connection. One can check that if the metric were introduced,
the boundary conditions set on the connection would simply require
the bulk metric to be spherically symmetric.

Viewed as a new example  (relative to inflow mechanism) 
of the brane-bulk interaction this work
has other interesting implementations. 
It shows how Einstein action on the brane can 
arise dynamically from higher-derivative terms in the bulk 
(for a similar result see also \cite{Zurab}).
The origin of this terms can be $\alpha'$ corrections of string theory.
The inclusion of such the Einstein term changes 
the problem of localization of gravity in brane-world scenarios.
The problem
is usually addressed in the following framework. The
brane is considered as a source to the gravity in the bulk.
By solving equations of motion in the bulk
one can find the background induced by the source.
Then the gravity on the brane is described as a normalizable zero mode of
the bulk fluctuations in this background.
The other possibility is to consider the theory on the brane
that includes the gravity \cite{Dvali}.
The attractive feature of this scenario is that localization
can be achieved even when the bulk theory is asymptotically flat. 
Besides, the short distance
behavior of the gravitational potential is modified, it becomes 
lower-dimensional.
Depending on the relative strength of bulk and brane gravity terms
there is an interesting switching between low and high dimensional
regimes ( see also \cite{Kirit}).
The mechanism we investigated here can provide an explanation to
how the gravity on the brane can be induced from bulk $\alpha'$
corrections.

\section*{Acknowledgment}

We are very grateful to Giulio Bonelli and Oleg Ruchayskiy for interesting 
discussions. B.K. would like to thank Olindo Corradini and Peter Langfelder.
A.B. thanks Danish research council.

\end{document}